\documentclass[12pt]{article}
\pdfoutput=1

\usepackage{epsfig,a4wide,amsmath,amssymb,amsfonts,mathrsfs}
\usepackage{mathtools}
\usepackage{color,yfonts}
\usepackage{graphicx}
\usepackage{subcaption}
\usepackage[utf8]{inputenc}

\numberwithin{equation}{section}

\newcommand{\be}{\begin{equation}}

\newcommand{\ee}{\end{equation}}
\numberwithin{equation}{section}
\newcommand{\mytitlefont}{\fontseries{mx}\selectfont}
\DeclareMathAlphabet{\titlemath}{OT1}{cmr}{mx}{n}

\begin{document}

\begin{titlepage}

\begin{center}

~\\[2cm]

{\fontsize{20pt}{0pt} \mytitlefont  Consequences of Analytic Boundary Conditions in AdS}

~\\[0.5cm]

{\fontsize{14pt}{0pt} Gary~T.~Horowitz and Diandian~Wang}

~\\[0.1cm]

\it{   Department of Physics, University of California, Santa Barbara, CA 93106}

~\\[0.05cm]

\end{center}

  \vspace{60pt}

\noindent

We investigate the effects of an analytic boundary metric for smooth asymptotically anti-de Sitter
 gravitational solutions. The boundary dynamics is then completely determined by the initial data due to corner conditions that all smooth solutions must obey. We perturb a number of familiar static solutions and explore the boundary dynamics that results. We
 find  evidence for a nonlinear asymptotic instability of the planar black hole in four and six dimensions. 
In four dimensions we find indications of at least exponential growth, while in six dimensions, it appears  that a singularity may form in finite time on the boundary. This instability extends to pure AdS (at least in the Poincare patch). 
For the class of perturbations we consider, there is no sign of this instability in five dimensions.

\vfill

    \noindent

  \end{titlepage}

   \newpage

\tableofcontents
\baselineskip=16pt
\section{Introduction}
Since the boundary of an asymptotically anti-de Sitter (AdS) spacetime is timelike, one must specify a (conformal) metric on this boundary as well as initial data to determine a solution to Einstein's equation. The solution will be smooth only if it satisfies an infinite set of compatibility conditions between the initial data and boundary data. This requirement is in addition to the usual constraint equations on the initial data, and takes the form of  corner conditions that must be satisfied on the co-dimension two surface where the initial data surface hits the  conformal boundary \cite{Friedrich:1995vb,Enciso:2014lwa,Carranza:2018wkp}. 

These corner conditions arise for the following reason. Given initial data for Einstein's equation, the evolution equations determine the second time derivative of the spatial metric. Higher time derivatives can be computed by taking derivatives of the evolution equations. The net result is that all time derivatives of the spatial metric can be computed at each point on the initial data surface. Taking the limit as the point goes to spatial infinity, one obtains all time derivatives of the boundary metric at the initial time.
To obtain a smooth solution in the bulk, the boundary metric must be compatible with these time derivatives. There is still freedom to choose a smooth (nonanalytic) boundary metric, but if the boundary metric is analytic, it is uniquely determined by the initial data through these corner conditions.

In this paper we explore the consequences of requiring that the boundary metric is analytic.  We investigate  what boundary dynamics is generated by perturbations of well known bulk solutions. For simplicity, we consider vacuum solutions to Einstein's equation (with a negative cosmological constant) in $D$ dimensions, and assume $D-2$ translation (and reflection) symmetries. We will compactify these directions into a torus $T^{D-2}$. In this case, there is a natural conformal frame for the boundary metric in which the size of one circle is kept fixed. It is also natural to let $t$ be proper time along the curves orthogonal to the torus.  For example, in $D=4$
the boundary metric can be put into the form:
\be\label{eq:bdy1}
ds^2|_{\text{bdy}} = -dt^2 + d\chi^2 + F(t) d\phi^2.
\ee
Since $F(t)$ measures the ratio of the size of two circles, it is conformally invariant.

We will consider static solutions and perturb the metric on a $t=0$ surface keeping the $(D-1)$-dimensional scalar curvature equal to $2\Lambda$, where $\Lambda$ is the cosmological constant. Setting the extrinsic curvature to zero, we satisfy the constraints and obtain time symmetric initial data for a nearby solution. We write the full time dependent evolution in  a convenient gauge and expand all metric functions in powers of $1/r$ (with time dependent coefficients), where $1/r^2$ is the conformal factor that results in the boundary metric (\ref{eq:bdy1}). We then solve Einstein's equation and its time derivatives order by order in $1/r$. This determines the time derivatives of all the coefficients in the expansion, evaluated at $t=0$.  In particular, we obtain time derivatives of $F(t)$. In principle, all time derivatives can be determined this way,  however in practice, we only compute a finite number of them. In several cases these time derivatives indicate that $F(t)$ grows rapidly suggesting an instability.

We start with the four-dimensional planar black hole, and add a perturbation that vanishes at infinity faster than $M/r$. We then use the corner conditions to compute the first 20 time derivatives of $F(t)$. This Taylor series indicates exponential growth of $F(t)$, so a small finite change in the initial data produces a huge change in the boundary dynamics. In fact, an $a_2/r^2$ perturbation of the black hole appears to result in $F(t) > e^{a_2 t^4}$. The fact that higher powers of $a_2$ appear in the time derivatives shows that this is a nonlinear instability.

\begin{figure}[t]
\begin{center}
\includegraphics[width=.45\textwidth]{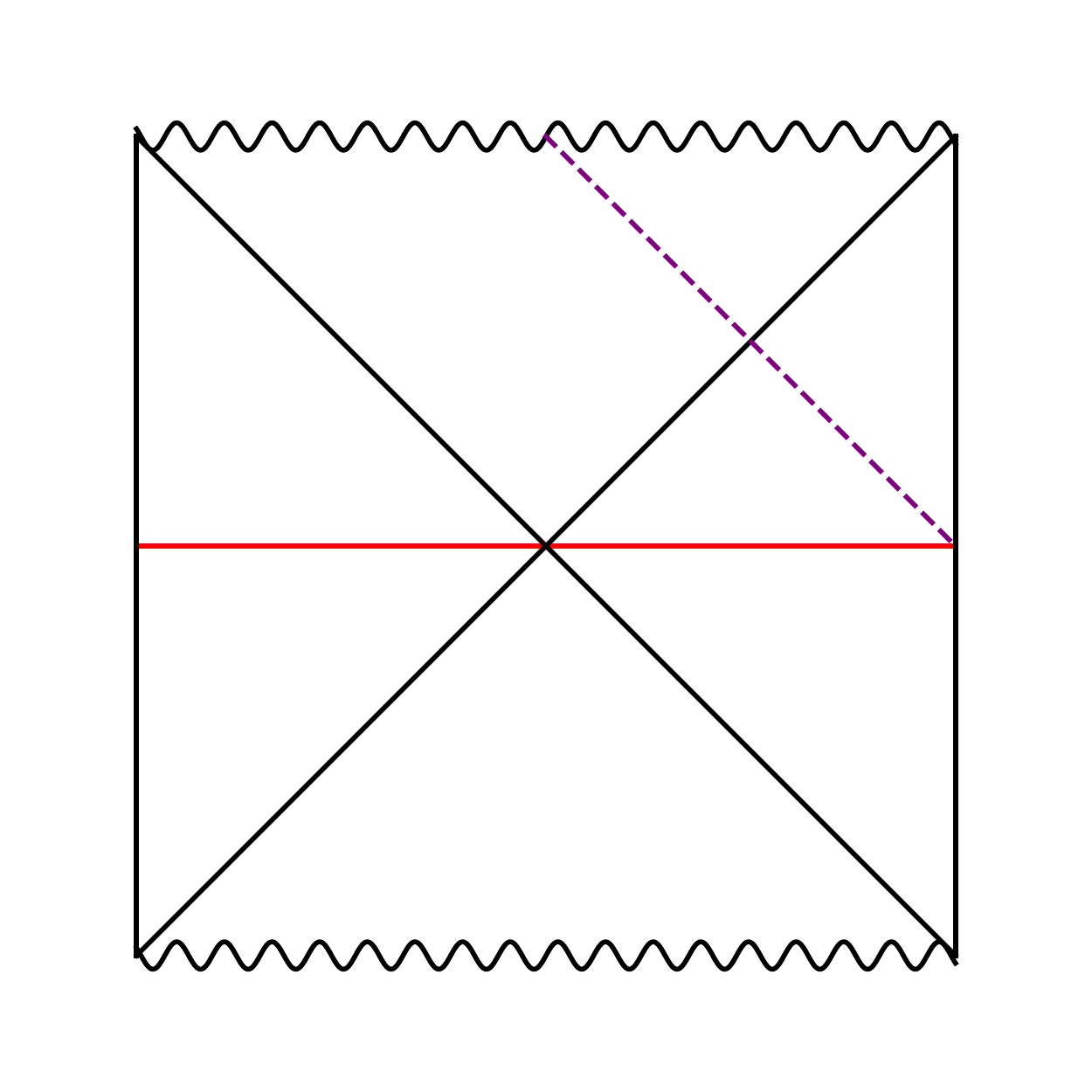}
 \caption{\label{fig:penrose_pert} Two different ways of perturbing the AdS black hole. The red horizontal line at $t=0$ is where our perturbed initial data will be, whereas the purple dashed line (an ingoing null surface) is where the perturbation in \cite{Dunn:2018xdm} lies.}
 \end{center}
\end{figure}

At first sight, this instability might seem to contradict a recent proof of the stability of the the four-dimensional planar AdS black hole  \cite{Dunn:2018xdm}. However this proof applies to perturbations on an ingoing null surface anchored at the boundary.  This characteristic data does not determine time derivatives of the boundary metric, so one does not have to impose an infinite set of corner conditions to obtain smooth solutions. The proof in  \cite{Dunn:2018xdm} assumes the usual static boundary metric. The difference is illustrated in Fig.\,\ref{fig:penrose_pert}.

This instability is very different from other instabilities that have been found for AdS black holes, 
 such as the superradiant instabilities of rotating black holes \cite{Cardoso:2004hs}. The latter  result from perturbations scattering off a black hole with increased amplitude and then bouncing off infinity and scattering off the black hole repeatedly.  It arises for any smooth boundary metric. The instability we discuss here requires an analytic boundary metric and can be found from a local calculation in a neighborhood of the corner where the initial data surface hits the boundary at infinity. It cannot arise in any spherically symmetric solution since there is no conformally invariant dynamics on the boundary. There is always a conformal frame in which the boundary metric is a static cylinder $R\times S^{D-2}$.

After investigating the four dimensional black hole, we then perturb 
other solutions in four and higher dimensions. For the six dimensional planar black hole, we find a similar instability. In fact, in this case there is evidence that the boundary metric becomes singular in finite time. We also extend the instability results to the Poincare patch of pure AdS (in both four and six dimensions). For the AdS soliton, the evidence for an instability is inconclusive. In five dimensions, we do not find an analogous instability for either the planar black hole or global AdS. In fact, in this case, the boundary seems to be unchanged by any power series perturbation.

In the next three sections we consider examples in four, five, and six dimensions respectively. In the last section we provide some insight into this instability and  mention some open problems.

\section{4D Examples}

\subsection{Planar black hole}

In this subsection, we investigate perturbations to the four-dimensional planar AdS black hole: 
\begin{align}
\label{eq:bh_unpert}
ds^2=  -\left(r^2 - \frac{M}{r}\right)dt^2 + \left(r^2 - \frac{M}{r}\right)^{-1} {dr^2}+ r^2 (d\chi^2 + d\phi^2).
\end{align}
 We have chosen to periodically identify $\chi$ and $\phi$ with periods $\Delta \chi$ and $\Delta \phi$ so the translationally invariant perturbations we will consider have finite energy, but the dynamical results are independent of this compactification. The boundary of this solution has topology $S^1 \times S^1 \times R$, or $T^2 \times R$. Taking a constant-$t$ slice gives a spacelike hypersurface with the metric given by the last three terms in (\ref{eq:bh_unpert}). We will perturb this initial data and study its time evolution. 

Consider a family of time-symmetric initial data of the form:
\begin{align}
\label{eq:4d_init}
ds^2|_{t=0} =\frac{dr^2}{\alpha(r) \beta(r)}+ \alpha(r) d\chi^2 + r^2 d\phi^2.
\end{align}
 Due to time symmetry, the extrinsic curvature is zero and the momentum constraint is automatically satisfied. The only non-trivial constraint is then the Hamiltonian constraint 
\begin{align}
\label{eq:ode_beta}
(2 \alpha+r \alpha^\prime) \beta^\prime + (4 \alpha^\prime +2 r \alpha^{\prime\prime}) \beta=12 r.
\end{align}
We can choose $\alpha(r)$ freely and solve this equation for $\beta(r)$. The planar black hole (\ref{eq:bh_unpert}) corresponds to $\alpha(r) = r^2$ and $\beta(r) = 1- M/r^3$. Since a small perturbation should not change the leading asymptotic behavior of (\ref{eq:bh_unpert}),  we consider
\begin{align}
\label{eq:alpha_a2}
\alpha(r) = r^2 + \frac{a_2}{r^2},
\end{align}
which fixes $\beta(r)$ to be
\begin{align}\label{eq:betaexp}
\beta(r) = \frac{e^{\frac{a_2}{4\, r^{4}}}}{r^{3}}\left[c_1 +\frac{3 \,a_2^{3 / 4} \,\Gamma\left(-\frac{3}{4}, \frac{a_2}{4\, r^{4}}\right)}{8 \sqrt{2}\, }\right]
\end{align}
up to the integration constant $c_1$, where $\Gamma(a,z)$ is the incomplete gamma function. 

As explained in the introduction, the corner conditions will be used to determine the boundary evolution. For this purpose, it is sufficient to express $\beta(r)$ as a power series near $r=\infty$. Since $\Gamma\left(-\frac{3}{4}, z\right) \propto z^{-3/4}$ for small $z$, we get 
\begin{align}
\beta(r) &= 1 +\frac{c_{1}+\frac{3 \,a_{2}^{3 / 4} \, \Gamma\left(-\frac{3}{4}\right)}{8 \sqrt{2}}}{r^{3}}+\frac{a_2}{r^{4}} + \cdots \\
&\equiv 1+ \frac{b_3}{r^3} + \frac{a_2}{r^4} + \cdots,
\end{align}
where we have defined a more convenient constant $b_3$, and $\Gamma(z)$ is the Euler gamma function. For this to be a ``subleading" perturbation to the black hole metric, we need to set $b_3 = -M$ so that the perturbed initial metric becomes
\begin{align}
\label{eq:pert_a2}
ds^2|_{t=0,\text{perturbed}} =\left(r^2-\frac{M}{r}+\frac{2a_2}{r^2}+\cdots\right)^{-1} dr^2+ \left(r^2 + \frac{a_2}{r^2}\right)d\chi^2 + r^2 d\phi^2.
\end{align}
 This change in the metric is bounded outside the event horizon, and vanishes as $a_2$ goes to zero. So it is a valid (nonlinear) perturbation to the black hole geometry. 

With this perturbed initial metric, we are ready to see what happens as it evolves. We will do this in a particular coordinate system, but the result is coordinate independent. We will use diffeomorphism invariance to (locally) put the metric in the form
\begin{align}\label{eq:genmetric}
ds^2 &= G(r,t) \,dt^2 +\frac{dr^2}{A(r,t)B(r,t)} +  A(r,t) \,d\chi^2 + r^2 F(t) \,d\phi^2,\\[5pt]
&F(0)=1, \quad \; A(r,0) = \alpha(r), \quad \; B(r,0) = \beta(r),
\end{align}
where $g_{rt}=0$ and $g_{\phi\phi}=r^2 F(t)$ are our gauge conditions. As explained in the introduction, we then expand $G(r,t)$, $A(r,t)$ and $B(r,t)$ as power series of $1/r$:
\begin{align}\label{eq:expcoef}
A(r,t)=r^{2}+\sum_{n=0}^{\infty}\frac{A_n(t)}{r^n},\quad
B(r,t)=\sum_{n=0}^{\infty}\frac{B_n(t)}{r^n},\quad
G(r,t)=-r^{2}+\sum_{n=0}^{\infty}\frac{G_n(t)}{r^n}.
\end{align}
These three functions describe the general metric with our symmetries. Setting $g_{\phi\phi}=r^2 F(t)$ instead of $g_{\phi\phi}=r^2$  allows us to remove the time dependence in the leading term in $A(r,t)$. Residual gauge freedom to reparametrize time on the boundary has been used to fix the leading term in  $G(r,t)$ to be  $-r^2$. Metric components proportional to $r$ cannot appear as a consequence of Einstein's equation evaluated near the boundary. We have used this fact preemptively. 

Rescaling (\ref{eq:genmetric})   by $1/r^2$  and using (\ref{eq:expcoef}) yields the boundary metric
\be
\label{eq:bdy}
ds^2|_{\text{bdy}} = -dt^2 + d\chi^2 + F(t) d\phi^2,
\ee
so $F(t)$ controls the boundary evolution. Notice that $B_0(t)$ is free in our ansatz, but Einstein's equation sets $B_0(t)=1$ for all $t$. This means $g_{rr}=1/r^2$ to leading order, so that $z=1/r$ to leading order, where $z$ is the standard Fefferman-Graham radial coordinate in which
\begin{align}
ds^2 = \frac{1}{z^2} \left[dz^2 +\left( \sum_{n=0}^{\infty}\gamma_{
\mu\nu}^{(n)}(x)z^n\right) dx^\mu dx^\nu\right].
\end{align}
This means that we can also obtain Eq.\,(\ref{eq:bdy}) by writing the metric in Fefferman-Graham coordinates and extracting $\gamma_{\mu\nu}^{(0)}$.

Imposing Einstein's equation and its time derivatives and solving them order by order in powers of $1/r$ allow us to obtain time derivatives of all the functions including $F(t)$. (For more details see \cite{Horowitz:2019dym}.) The first 20 time derivatives of $F(t)$ have been calculated:
\begin{align}\label{eq:Fexp}
F(t) =& \;1+\frac{1}{3} a_2 t^{4}+\frac{11}{105} a_2^{2} t^{8}+\frac{73 }{1260}M^{2} a_2 t^{10}+\frac{21482 }{467775}a_2^{3} t^{12}+\frac{1887 }{21560} M^{2} a_2^{2} t^{14}\nonumber\\[10pt]
&+\frac{\left(39541905  M^{4} a_2+70796224 a_2^{4}\right) t^{16}}{2724321600} +\frac{437831897 }{4086482400}M^{2} a_2^{3} t^{18}\nonumber \\[10pt]
&+\frac{\left(29268111459375 M^{4} a_2^{2}+10234594085504 a_2^{5}\right) t^{20}}{593970216840000} +\mathcal{O}(t^{22})\\[10pt]
\approx &\; 1+0.33\, a_2 t^{4}+0.10 \,a_2^{2} t^{8}+ 0.058\,M^{2} a_2 t^{10}+0.046\, a_2^{3} t^{12}+0.088\, M^{2} a_2^{2} t^{14} \nonumber\\[10pt]
&+\left(0.015\, M^{4} a_2+0.026\, a_2^{4}\right) t^{16} +0.11\,M^{2} a_2^{3} t^{18} \nonumber\\[10pt]
&+\left(0.049\, M^{4} a_2^{2}+ 0.017\,a_2^{5}\right) t^{20} +\mathcal{O}(t^{22}).
\end{align}
Only even powers of $t$ appear since our initial data was time symmetric. The structure of the terms can be understood from a simple scaling argument  \cite{Horowitz:2019dym}. Notice that the metric (\ref{eq:genmetric})  is invariant under
\begin{align}\label{eq:scaling}
 r= \lambda \tilde r, \ \ (t,\chi,\phi) = (\tilde t,\tilde \chi, \tilde\phi) / \lambda,\ \  (A,G) = \lambda^2 (\tilde A,  \tilde G),\ \  (B,F) = (\tilde B,\tilde F).
\end{align} 
If we define the dimension of a quantity to be the power of $\lambda$ that it acquires under this transformation, then $M$ has dimension three and $a_2$ has dimension four.  Noting that $t$ has dimension $-1$, each term in (\ref{eq:Fexp}) has dimension zero as required for $F$.

A remarkable feature of this expansion is that all the coefficients are positive (for $a_2 > 0$). So we can get a lower bound to the growth of $F(t)$ by focussing on the terms that are independent of $M$. These terms can be written as a  power series in $T\equiv a_2 t^4$. To compare with $e^T$, we compute derivatives of $F$ with respect to $T$ at $T=0$ (to 2 decimal places) and find: 
\begin{align}
\begin{split}
F(T=0)=1.00,\qquad
F^{(1)}(T)|_{T=0}&=0.33, \\
F^{(2)}(T)|_{T=0}=0.21, \qquad
F^{(3)}(T)|_{T=0}&=0.28, \\
F^{(4)}(T)|_{T=0}=0.62, \qquad
F^{(5)}(T)|_{T=0}&=2.07.
\end{split}
\end{align}

\begin{figure}[t]
\begin{center}
\includegraphics[width=.6\textwidth]{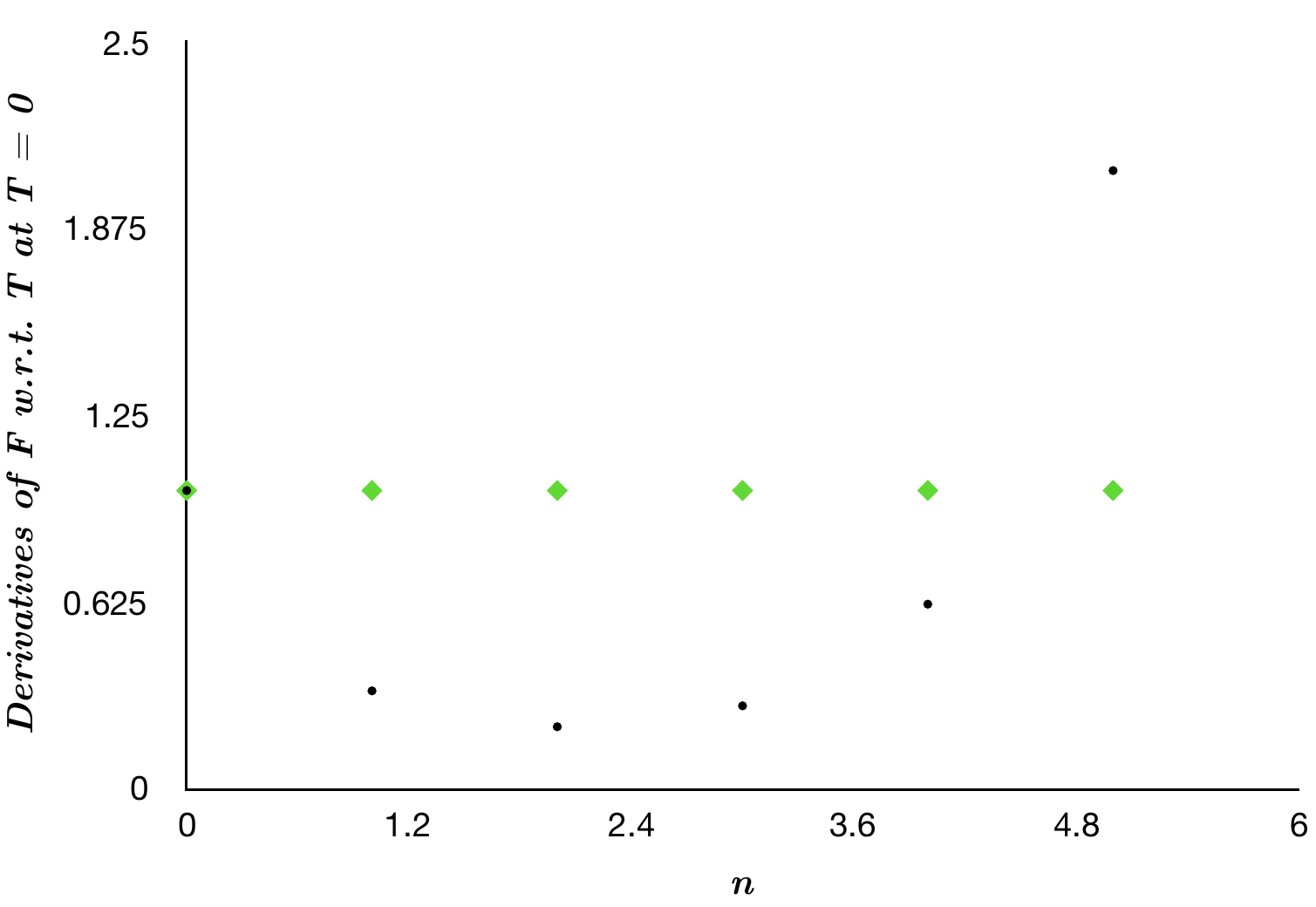}
 \caption{\label{fig:Fn_4dbh_a2_t4} Mass-independent contributions to $F^{(n)}(T)|_{T=0}$ plotted against $n$ for the perturbation given by Eq.\,(\ref{eq:alpha_a2}) (black dots) and for a reference function $F_{\text{ref}}(T)=e^T$ (green diamonds). Since the mass-dependent terms are all positive, this gives a lower bound on the growth of $F(T)$ if the pattern displayed in the figure continues. This suggests that the function will grow at least at fast as $e^{a_2 t^4}$.}
 \end{center}
\end{figure}

We can plot $F^{(n)}(T)|_{T=0}$ against $n$. Fig.\,\ref{fig:Fn_4dbh_a2_t4} shows how it compares to the function $e^T$. Although the first few derivatives are less than one, the last one we have computed is larger than one and they appear to be increasing,  suggesting that $F(t)$ will grow at least like $e^{a_2t^4}$. If we compute derivatives of $F$ with respect to $\tilde T = \sqrt {a_2} t^2$, we get:
\begin{align} \label{t2growth}
\begin{split}
F(\tilde T=0)=1.00,\qquad
&F^{(2)}(\tilde T)|_{\tilde T=0}=0.66, \\
F^{(4)}(\tilde T)|_{\tilde T=0}=2.51, \qquad
&F^{(6)}(\tilde T)|_{\tilde T=0}= 33.06, \\
F^{(8)}(\tilde T)|_{\tilde T=0}=1048, \qquad
&F^{(10)}(\tilde T)|_{\tilde T=0}=62527.
\end{split}
\end{align}
clearly showing that $F(t)$ is growing much faster than $\cosh(\sqrt{ a_2} t^2)$. This is a nonlinear instability since the higher powers of $a_2$ only arise due to nonlinearities in Einstein's equation.

We now consider perturbations that fall off faster than $a_2/r^2$.  The above dimension counting argument shows that adding a perturbation $a_3/r^3$ to $\alpha(r)$ will induce powers of $a_3 t^5$ in the expansion of $F$. But since only even powers of $t$ can appear in the evolution of time symmetric data,  only two terms will be nonzero in the first 20 time derivatives. So we will instead add an $a_4/r^4$ term next. However, after that we will  turn on multiple $a_i$'s, including odd $i$. These coefficients can combine to give many non-zero cross-terms in the expansion of $F(t)$, and we will examine their signs. Before doing that, we turn on a single $a_i$ at a time.

Replacing Eq.\,(\ref{eq:alpha_a2}) with
\begin{align}
\alpha(r) = r^2 + \frac{a_4}{r^4},
\end{align}
and solving the constraint (\ref{eq:4d_init}), we obtain
\begin{align}
\beta(r)=c_1 \frac{r^{12}}{\left(2 r^{6}-a_4\right)^{5 / 2}}+\frac{4 r^{12}\, {}_2F_1\left(2, \frac{5}{2}; \frac{7}{2}; 1-\frac{2 r^{6}}{a_4}\right)}{5 a_4^{2}},
\end{align}
where ${}_2 F_1(a,b;c;z)$ is the hypergeometric function.  We again expand this in a power series in $1/r$:
\begin{align}
\beta(r) = 1+\frac{-\frac{3 \pi\sqrt{a_4} }{4 \sqrt{2}}+\frac{c_{1}}{4 \sqrt{2}}}{r^{3}}+\frac{2 a_4}{r^{6}}+\cdots
\equiv 1+\frac{b_3}{r^3} + \frac{2 a_4}{r^6} + \cdots.
\end{align}
So, setting $b_3 = -M$ as before, our initial data describes another perturbation of the black hole.

Using the same coordinate system and notation as before, we again compute the first 20 time derivatives of $F(t)$: 
\begin{align}
\begin{split}
F(t) = \,&1+\frac{a_{4} t^{6}}{5}+\frac{\left(4915 M^{2} a_{4}+4728 a_{4}^{2}\right) t^{12}}{46200} \\[5pt]
&+\frac{\left(24593045 M^{4} a_{4}+160990052 M^{2} a_{4}^{2}+55433856 a_{4}^{3}\right) t^{18}}{571771200} \\[10pt]
=\,& 1 + 0.20 a_4 t^6 + \left(0.11 M^2 a_4 + 0.10 a_4^2\right) t^{12} \\[5pt]
&+ \left(0.043 M^4 a_4 + 0.28 M^2 a_4^2 + 0.097 a_4^3\right) t^{18} + \mathcal{O}(t^{22}).
\end{split}
\end{align}
Notice that all the terms are again positive, and the terms independent of $M$ appear in the dimensionless combination $T = a_4 t^6$. The first few derivatives of $F$ with respect to $T$ can be computed and are slightly less than one. Even though these derivatives are less than one, they are growing and likely to exceed one. If so, $F$ will grow faster than $e^{a_4 t^6}$. If one computes derivatives with respect to  $\tau \equiv a_4^{1/3} t^2$ and focuses on terms containing only powers of $a_4$, we have
\begin{align}
\begin{split}
F(\tau)|_{\tau=0}=1.00,\qquad
F^{(3)}(\tau)|_{\tau=0}&=1.20, \\
F^{(6)}(\tau)|_{\tau=0}=73.68, \qquad
F^{(9)}(\tau)|_{\tau=0}&=35181.6,
\end{split}
\end{align}
clearly showing a rapid growth.

To summarize, for both $a_2/r^2$ and $a_4/r^4$ perturbations in $\alpha(r)$ to the planar black hole, we find a rapid growth of the size of one circle relative to the other in the boundary metric (and in the asymptotic region). 

More generally, we have also studied perturbations given by
\be
\label{eq:4d_alpha_gen}
\alpha(r) = r^2 + \sum_{n=1}^{9} \left(\frac{a_{2n}}{r^{2n}}+\frac{a_{2n+1}}{r^{2n+1}}\right)
\ee
and computed the first twenty time derivatives of $F(t)$. (We could  have included higher order terms in this expansion but they would not contribute to the first 20 derivatives.) Remarkably, {\bf for positive $a_{2n}$ and negative $a_{2n+1}$, all 133 terms in $F(t)$ are positive} including cross-terms involving two or more $a_i$'s.  There is thus a large class of unstable perturbations.

\subsection{Pure AdS: Poincare patch}

To investigate perturbations to the Poincare patch of pure AdS, we can simply take our results for the planar AdS black hole and set $M$ to zero. However, there is an important difference. A pivotal feature of the black hole is the existence of a horizon. We have used this fact to turn on $a_i$ individually while keeping the perturbation small outside the horizon. Indeed, a perturbation like $1/r^n$ for positive $n$ blows up near the origin. This was not a problem for the black hole, but a neighborhood of $r=0$ is now part of our initial data. If we want our perturbation to die off at the origin, we will have to choose it more carefully. Since our instability only depends on the asymptotic behavior of the initial data, one could modify the initial data by hand in the interior in a smooth but not analytic way so that it vanishes near the Poincare horizon. But since the boundary is required to be analytic, one might ask if there is an analytic perturbation that is unstable.

For an analytic perturbation to the initial data, keeping all $a_i$ positive will obviously make $\alpha(r)$ diverge at small $r
$. Making all $a_i$ negative, on the other hand, will give $\alpha(r)=0$ for some finite $r$. Therefore, a finite perturbation everywhere necessitates a mixture of positive and negative $a_i$ terms in the asymptotic expansion. Fortunately, for our generic perturbation (\ref{eq:4d_alpha_gen}), having all terms of $F(t)$ being positive precisely requires a mixture of two signs. Although a growing $F(t)$ does not by itself suggest an instability, we have seen that having $a_2/r^2$ and $a_4/r^4$ terms already leads to growth faster than exponential, and other terms will only enhance this instability. Finite functions having positive $a_{2n}$ and negative $a_{2n+1}$ are easy to construct. One simple example is
\begin{align}
\alpha(r) = r^2 +  \frac{Ae^{-1/r}}{r^2},
\end{align}
for a small parameter $A$. This shows that Poincare patch of AdS is unstable under an analytic perturbation with analytic boundary condition.

\subsection{Soliton}
\label{sec:4d_soliton}
In this subsection we consider the AdS soliton \cite{Witten:1998zw,Horowitz:1998ha} which is expected to have the lowest energy among solutions with 
$T^2 \times R$ boundary topology. This has been  proven recently for time symmetric initial data with U(1)$^{D-2}$ symmetry \cite{Barzegar:2019vaj}, which is just what we are considering. Of course having minimum energy does not preclude instabilities associated with adding small finite perturbations. So we now ask whether this ground state exhibits the same type of instability as the black hole. 

The metric is a double analytic continuation of that of the planar black hole:
\begin{align}
\label{eq:soliton}
ds^2=  - r^2 dt^2 +  \left(r^2 - \frac{M}{r}\right)^{-1} {dr^2} +
\left(r^2 - \frac{M}{r}\right)d\chi^2 + r^2  d\phi^2.
\end{align}
 The effect of perturbing this solution was carried out in \cite{Horowitz:2019dym} to 12 orders in $t$. Here we extend the calculation to higher orders.

The calculation for the soliton is almost exactly the same as that for the black hole. 
In fact, the results for both can be obtained in one single calculation,  with 
\be
\alpha(r) = r^2 + \frac{a_1}{r} + \frac{a_2}{r^2}, \qquad \beta(r) = 1 + \frac{b_3}{r^3} + \cdots,
\ee
setting $a_1=0$, $b_3=-M$ for the black hole and $a_1 = -M$, $b_3=0$ for the soliton only after the computation is complete.  Note that the terms containing only powers of $a_2$ would be exactly the same as before, because these terms can be obtained by setting $M=0$ in both cases so that the two problems become identical. The terms containing powers of $M$ will however be different, as $M$ arises from different parts of the metric in each case. The first 16 terms of the Taylor series for $F(t)$ are:
\begin{align}
F(t) = \,&1+\frac{1}{3} a_{2} t^{4}+\frac{11}{105} a_{2}^{2} t^{8}+\frac{11 }{2520}M^{2} a_{2} t^{10}+\frac{21482 }{467775}a_{2}^{3} t^{12} \\[10pt]
&+\frac{102937 }{20180160}M^{2} a_{2}^{2} t^{14}
+\frac{\left(-39555 M^{4} a_{2}+17699056 a_{2}^{4}\right) t^{16}}{681080400} + \mathcal{O}(t^{18}).
\end{align}
Notice the minus sign in the $M$ dependent coefficient of $t^{16}$.
Since  $M$ is finite and fixed while $a_2$ should be taken arbitrarily small to be considered a perturbation, the order $t^{16}$ term is negative overall. Furthermore, notice that whatever sign we choose for $a_2$, this expansion will contain both positive and negative terms. Thus, we cannot conclude anything about the growth of $F(t)$ from this expansion. An $a_4/r^4$ perturbation also gives negative terms.  Therefore, at least for $a_2$ and $a_4$ perturbations, we cannot conclude that there is an instability.

We now mention a subtlety for the soliton that is not present in the case of black holes. For the unperturbed soliton to have no conical singularity where $\alpha(r)=0$, we need to choose the correct period for the $\chi$ circle. However, with the correct period chosen for the unperturbed soliton, a generic perturbation will lead to a conical singularity where the perturbed $\alpha$ equals zero. As the change in the metric will be small, a small change in the period of the $\chi$ circle can be employed to restore smoothness. We always include the necessary change in the period as part of our definition of the perturbation.

\section{5D Examples}

\subsection{Planar black hole}
Given evidence for an instability in four dimensions, a natural question to ask is what happens in other dimensions. In three dimensions, the boundary metric is always conformal to a static cylinder, so there is no analogous instability. We therefore consider higher dimensions. We will first look at the five-dimensional case  and then move on to six dimensions in the next section.

The unperturbed metric for a 5D planar black hole in AdS is given by
\begin{align}\label{eq:5dbh}
ds^2=  -\left(r^2 - \frac{M}{r^2}\right)dt^2 + \left(r^2 - \frac{M}{r^2}\right)^{-1} dr^2 + r^2 (d\chi^2 + d\phi^2 + d\psi^2).
\end{align}
Now we have three circles, so we can perturb the solution in several ways. One option is to perturb one of the circles as before by adding subleading terms to the metric, while keeping the $T^2$ symmetry for the remaining two circles. In other words, we start with the perturbed initial data
\begin{equation}
\begin{gathered}
ds^2|_{t=0} =\frac{dr^2}{\alpha(r) \beta(r)}+ \alpha(r) \,d\chi^2 + r^2 (d\phi^2+d\psi^2), \\
\alpha(r)= r^2 +\sum_{n=3}^{\infty} \frac{a_n}{r^n},
\end{gathered}
\end{equation}
where each $a_n$ can be independently zero or non-zero. Note that we start from $n=3$ because the mass term is of order $1/r^2$. Now $\beta(r)$ is again obtained by solving the initial data constraint, and the integration constant is chosen to match the mass term to that of the unperturbed metric. To obtain the time dependence, we again write the metric in the form
\begin{align}
ds^2 = G(r,t) \,dt^2 +\frac{dr^2}{A(r,t)B(r,t)} +  A(r,t) \,d\chi^2 + r^2 F(t) (d\phi^2+d\psi^2),
\end{align}
and again compute time derivatives of $F$. We find that they all vanish, i.e.,
\begin{align}
F(t) = 1 + \mathcal{O}(t^{14}).
\end{align}
This is strong evidence that $F=1$ to all orders, so our perturbation does not induce any time dependence on the boundary in five dimensions.

Alternatively, we can perturb a 2-torus and study its effect on the third circle. A $T^2$ symmetric perturbation can be written as 
\begin{equation}
\begin{gathered}
ds^2|_{t=0} =\frac{dr^2}{\alpha(r) \beta(r)}+ \alpha(r) (d\chi^2 +d\phi^2)+ r^2 \,d\psi^2, \\
\alpha(r)= r^2 +\sum_{n=3}^{\infty} \frac{a_n}{r^n},
\end{gathered}
\end{equation}
which evolves to (by choosing the gauge similarly)
\begin{align}
ds^2 = G(r,t) \,dt^2 +\frac{dr^2}{A(r,t)B(r,t)} +  A(r,t) (d\chi^2+d\phi^2) + r^2 F(t) \,d\psi^2.
\end{align}
Again, we find that the circle does not grow: $F(t) = 1 + \mathcal{O}(t^{14})$. Breaking this $T^2$ symmetry and choosing two independent perturbations for two circles results in the same answer, $F(t) = 1 + \mathcal{O}(t^{8})$, where higher orders have not been calculated.

It thus seems that five dimensions is very different from four dimensions, at least in the case of planar black holes. Adding perturbations that are powers of $1/r$ do not affect the boundary at all. Since our choice for the perturbation appears quite generic, it may be tempting to conclude that no perturbation can affect the boundary in 5D. However, the Fefferman-Graham expansion of solutions in 5D contains logarithmic terms which are not present in even dimensions. One might thus want to include logarithmic terms in the initial data perturbation and see whether they induce a time dependence on the boundary. We will give an argument in Sec.\,5 that they do.

%Pure AdS${}_5$ in the Poincare patch would be trivial because setting $M=0$ does not change the fact that our boundary metric seems to remain static. 

\subsection{Pure AdS: global}

We next consider perturbations of global $\text{AdS}_5$.\footnote{We cannot do a similar analysis of  global $\text{AdS}_4$ without breaking the spherical symmetry and keeping only a single Killing field. In five dimensions we can keep enhanced symmetry even after breaking spherical symmetry.}
The unperturbed global $\text{AdS}_5$ solution can be written using the following metric,
\begin{equation}
\begin{gathered}
ds^2 = -g(r)dt^2 + dr^2/a(r) + h(r) (d\theta^2 + \sin^2 \theta \,d\phi^2) + f(r) (d\chi + \cos\theta\, d\phi)^2,\\
g(r) = a(r) = r^2 + 1, \\
h(r) = f(r) = r^2/4.
\end{gathered}
\end{equation}
Now we perturb the initial data (restricting to time-symmetric initial data as before) so that
\begin{align}\label{eq:pert5D}
\begin{split}
f(r) &= r^2/4 \text{\,(unchanged)},\\
h(r) &= r^2/4 + \frac{A\, r^4}{B + r^6},
\end{split}
\end{align}
    which determines $a(r)$ through a first-order ODE (the initial data constraint ${}^4R = 2\Lambda$), which we solve numerically. Expanding in powers of $1/r$ yields
\begin{align}
h(r) = \frac{r^2}{4} + \frac{A}{r^2} - \frac{AB}{r^8}+\frac{AB^2}{r^{14}}-\frac{AB^3}{r^{20}}+\mathcal{O}(1/r^{26}).
\end{align}
%This allows us to make all coefficients positive if we want to, and can change the relative size of the coefficients by varying $B$. We will keep $B$ free in our calculation for corner conditions. 

\begin{figure}[t]
    \centering
    \includegraphics[width=0.6\textwidth]{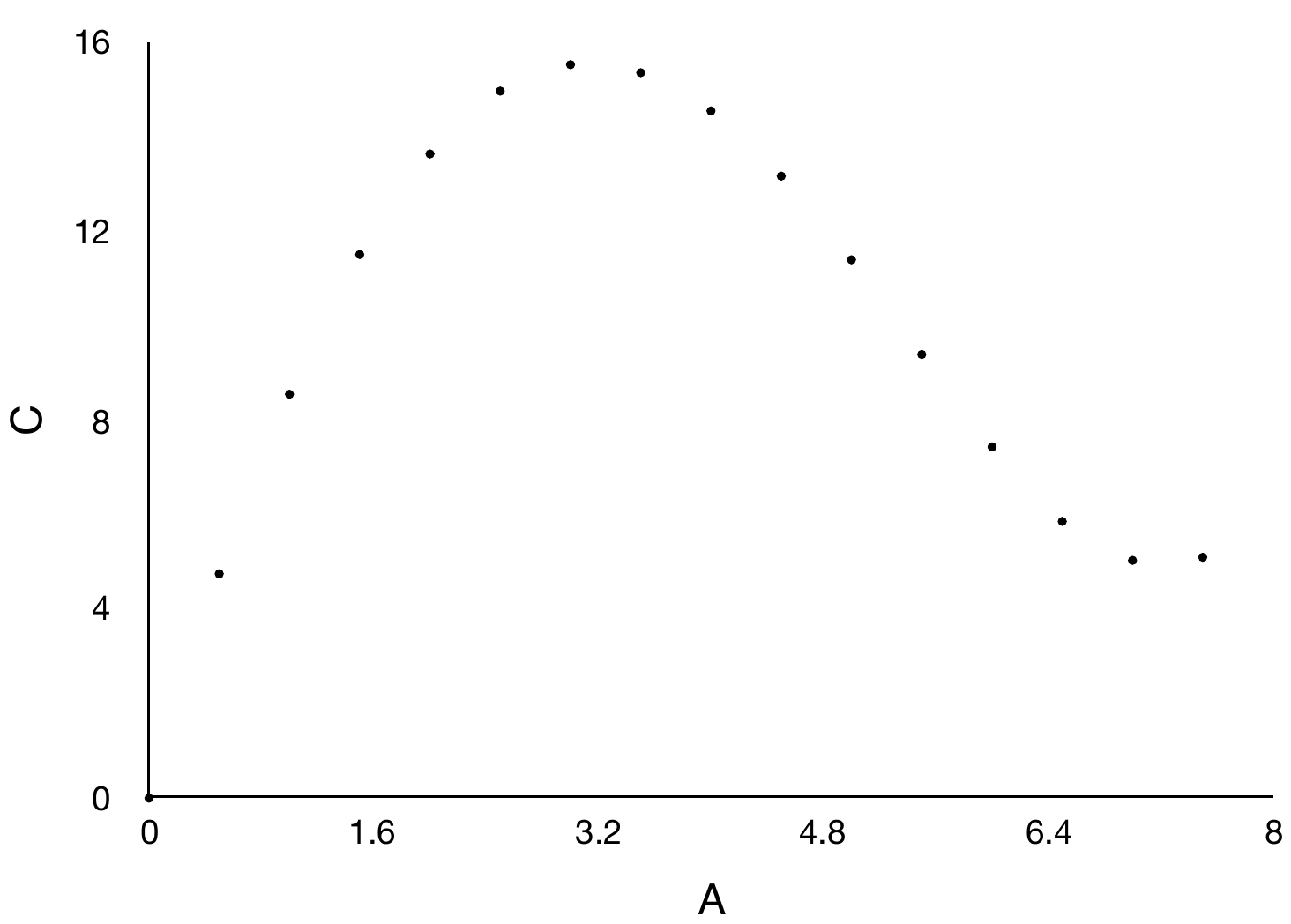}
    \caption{Relation between the free parameter, $A$, in the perturbation (\ref{eq:pert5D}) with $B=1$, and the constant, $C$, in the asymptotic expansion of $a(r)$ (\ref{eq:5dglobal_asymp}).}
    \label{fig:CvsA}
\end{figure}

For small $r$, using regularity at the origin to fix the integration constant, 
\begin{align}
a(r) = 1+\left(1-\frac{8A}{B}\right)r^2 + \mathcal{O}(r^4),
\end{align}
which, upon solving the ODE, gives the integration constant, $C$, in the large-$r$ expansion
\begin{align}
\label{eq:5dglobal_asymp}
a(r) = r^2 + 1+ \frac{C}{r^2} + \mathcal{O}(1/r^4).
\end{align}
Figure \ref{fig:CvsA} shows the numerically determined relation between $A$ and $C$, where $B$ is fixed to 1. 

Now for the evolution,  we use a gauge similar to the previous ones, and write
\begin{equation}
\begin{aligned}
ds^2 =\, G(r,t) dt^2 + \frac{dr^2}{A(r,t)} + H(r,t)&(d\theta^2+\sin^2\theta d\phi^2) + \frac{r^2}{4} F(t) (d\chi^2+\cos\theta d\phi)^2,\\
G(r,t) &= -r^2 + \sum_{n=0}^{\infty}\frac{G_n(t)}{r^n}, \\
A(r,t) &= A_b(t) r^2 + \sum_{n=0}^{\infty}\frac{A_n(t)}{r^n}, \\
H(r,t) &= \frac{r^2}{4}+ \sum_{n=0}^{\infty}\frac{H_n(t)}{r^n}.
\end{aligned}
\end{equation}
The boundary evolution of the function $A(r,t)$, i.e., the function $A_b(t)$ is found to be
\begin{align}
A_b(t)= 1 + \mathcal{O}(t^{12}),
\end{align}
so, as before, we can use the usual Fefferman-Graham radial coordinate to determine the boundary metric. This is equivalent to using $1/r^2$ as the conformal rescaling factor to obtain the conformal metric at the boundary. The only possible change in the boundary metric comes from $F(t)$. It is given by
\begin{align}
F(t) =  1 + \mathcal{O}(t^{12}).
\end{align}
So, up to this order, the corner conditions predict a static boundary metric if it is analytic. This is just like the 5D black hole. Although we have chosen a specific form of perturbation (\ref{eq:pert5D}), it is likely that the boundary would remain static under more general power series perturbations. Again, this should change if log terms are included. See Discussion section for an argument.

\section{6D Examples}
\subsection{Planar black hole}

The six dimensional planar black hole is similar to (\ref{eq:5dbh}) with the $T^3$ replaced by $T^4$ and $M/r^2$ replaced by $M/r^3$.  In analogy to the 5D case, we can (a) perturb one circle and calculate the response of the remaining $T^3$,
  (b) perturb a symmetric $T^2$  and calculate the response of the other symmetric $T^2$, or (c) perturb a symmetric $T^3$ and calculate the response of the remaining circle. In each case, we choose an ansatz similar to those of the previous examples and consider a perturbation of the form
\be
\alpha(r) = r^2 + \frac{a_{4}}{r^{4}} +\frac{a_{6}}{r^{6}}.
\ee
For case (a), we have
\begin{align}
F(t) = 1-\frac{a_4  t^6}{5}  -\frac{3 a_6  t^{8} }{35} - \frac{a_4^2 t^{12}}{2310}  - \frac{1791 a_4 a_6 t^{14}}{175175}  - \left( \frac{3169 M^2 a_4}{420420}  + \frac{103 a_6^2}{15925} \right)t^{16} + \cdots;
\end{align}
For case (b), we have
\begin{align}
F(t) = 1-\frac{a_4  t^6}{5}  -\frac{3 a_6  t^{8} }{35} - \frac{109 a_4^2 t^{12}}{5775}  - \frac{7543 a_4 a_6 t^{14}}{175175}  - \left( \frac{3169 M^2 a_4}{420420}  + \frac{67 a_6^2}{3185} \right)t^{16} +\cdots;
\end{align}
For case (c), we have
\begin{align}
F(t) = 1-\frac{a_4  t^6}{5}  -\frac{3 a_6  t^{8} }{35} - \frac{431 a_4^2 t^{12}}{11550}  - \frac{2659 a_4 a_6 t^{14}}{35035}  - \left( \frac{3169 M^2 a_4}{420420}  + \frac{81 a_6^2}{2275} \right)t^{16} +\cdots.
\end{align}
In each case, the prefactors of $a_4^2 t^{12}$ and $a_6^2 t^{16}$ are now both negative, and as a result these terms contribute negatively to the evolution regardless of the signs of $a_4$ and $a_6$.  This is quite different from the four dimensional case: for these two perturbations we do not see evidence for an exponential growth. 

However a shrinking circle can also indicate an instability since it is only the ratio of the size of the two circles that is conformally invariant.  To explore this, we turn on a more general perturbation given by
\be
\label{eq:6d_alpha_gen}
\alpha(r) = r^2 + \sum_{n=2}^{7} \left(\frac{a_{2n}}{r^{2n}}+\frac{a_{2n+1}}{r^{2n+1}}\right),
\ee
where $n$ starts at 2 because we want perturbations to be more subleading than the mass term (order $1/r^3$). Surprisingly, after computing 16 derivatives of $F(t)$,  all time dependent terms (there are 16 of them) are \textit{negative} for positive $a_{2n}$ and negative $a_{2n+1}$. This occurs in all three cases (a,b,c) above. This  suggests that for this class of perturbations $F(t)$ will monotonically decrease to zero. When $F=0$, the boundary metric develops a singularity where (a) the 3-torus, (b) the 2-torus, or (c) the circle shrinks to a point. 

If $F$ indeed vanishes in finite time, we could choose another conformal frame in which the shrinking torus is kept fix but the remaining circles expand. In this frame, the evolution would look very similar to the four dimensional case, however we would know that the expanding circles actually diverge in finite time. That might also be the case for the four dimensional solutions if the Taylor series fails to converge at a finite time.

\subsection{Pure AdS: Poincare patch}
As before, we can simply set $M=0$ and obtain results for the Poincare patch of AdS${}_6$. Just like the Poincare patch of AdS${}_4$, we will need to choose perturbations that die off near $r=0$. This requires mixed signs in the series expansion of $\alpha(r)$. In 6D, we now have the statement that all time dependent terms are negative for positive $a_{2n}$ and negative $a_{2n+1}$ (up to the order we have calculated). We can for example choose a function like
\begin{align}
\alpha(r) = r^2 + \frac{Br^{-1/r}}{r^4},
\end{align}
where $B$ is a small parameter indicating the size of our perturbation. This would give a singularity when $F(t)=0$ as in the  black hole example above. 

\subsection{Soliton}
Like in Sec.\,\ref{sec:4d_soliton}, we can obtain the results for the soliton easily. Adding $-M/r^3$ to Eq.\,(\ref{eq:6d_alpha_gen}) and setting $b_5=0$, we obtain $F(t)$ for the soliton in each of the three cases. Surprisingly, in all cases (a,b,c), we again find that all time dependent terms are negative for $a_{2n}>0$ and $a_{2n+1}<0$. However the term with the wrong sign in four dimensions was proportional to $M^4 a_2$, and the analogous term in six dimensions would be $M^4 a_4$. Since $M$ now has dimension five under the scaling (\ref{eq:scaling}),
 this has dimension 26 and would not show up until we reach $t^{26}$ terms in the Taylor expansion. Thus we should probably not conclude anything about the stability of the AdS soliton in six dimensions.

\section{Discussion}

We have studied small but finite perturbations of several static asymptotically AdS solutions of Einstein's equation. Assuming that the boundary metric is analytic, it is completely determined by the perturbed initial data due to corner conditions that all smooth solutions must obey. In five dimensions, the boundary metric remained static, but in both four and six dimensions, the  boundary metric becomes dynamical. For the planar black hole, we have found evidence for an asymptotic instability, where the size of one circle grows rapidly relative to the other. In six dimensions, it appears that this can lead to a curvature singularity in finite time. (This might also be true in four dimensions.) This instability extends to the Poincare patch of pure AdS in both four and six dimensions. It is not clear if it also applies to the AdS soliton in these dimensions.

How can we understand this instability? The linear terms in our expansions for the time dependence of the boundary metric have a simple explanation. Linear metric perturbations act like a massless scalar field. So consider the Poincare patch in $D$ dimensions
\be
ds^2 = \frac{1}{z^2} \left[ -dt^2 + dx_i dx^i + dz^2 \right],
\ee
where $i = 1, \cdots, D-2$.
Translationally invariant solutions of $\nabla^2 \Phi =0$ satisfy
\be
\ddot\Phi = \Phi'' + \frac{2-D}{z} \Phi' .
\ee
So if we expand
\be\label{eq:cexp}
\Phi = \sum_{n=0}^\infty c_n(t) z^n,
\ee
the coefficients must satisfy
\be\label{eq:recursion} 
\ddot c_{n-2} = n(1+n - D) c_{n}.
\ee
The familiar modes of a massless scalar in AdS${}_D$  are proportional to $z^{D-1}$ asymptotically. Setting $n = D-1$ in (\ref{eq:recursion}) we see that this does not trigger any time dependence in the other terms, and it is consistent to keep $\Phi = 0$ on the boundary $z=0$. However, if higher order terms are nonzero at $t=0$, they will trigger time dependence in the lower order terms. If this continues down to $c_0$, then $\Phi(z=0)$ becomes time dependent. 
This does not happen for odd  $D$, since  $D-1$ is even and  (\ref{eq:recursion}) relates coefficients differing by two, so even if $c_{D-1}$ becomes time dependent,  there is no time dependence in the lower order  even terms.

For even $D$, the situation is very different. Now, if $c_n\ne 0$ for any even integer $n$ larger than $D-1$, Eq.  (\ref{eq:recursion}) implies time dependence in all lower order even coefficients including $c_0$. In $D=4$ for example,  $c_4\ne 0$ implies $c_0 = -c_4 t^4/3$, and $c_6 \ne 0$ implies $c_0 = - c_6 t^6/5$. For $D=6$, $c_6 \ne 0 $ implies $c_0 =c_6 t^6/5$, and $c_8 \ne 0 $ implies $c_0 = 3 c_8 t^8 /35$.  These coefficients agree exactly with the linear terms in the expansions for $F(t)$ with the translation
$a_n = - c_{n+2}$. (The sign is unimportant for a linear perturbation.)

Returning to odd dimensions, consider adding a term $\tilde c_{D-1} z^{D-1} \ln z$ to the expansion (\ref{eq:cexp}). Then one finds $\ddot c_{D-3} = (D-1) \tilde c_{D-1}$. This implies a nonzero $\ddot c_0$. Hence log terms in the perturbation will generate time dependence on the boundary in odd dimensions.

Given this simple explanation for why a power law fall off in $z$ (or $1/r$) produces a power law growth in $t$ on the boundary, the surprise lies in the fact that the nonlinearities of general relativity appear to enhance this to (at least) an exponential growth. We stress that only the asymptotic initial data (and all its derivatives) are needed to generate this instability. 

Physically, one can avoid this asymptotic instability by either not using an analytic boundary metric or allowing bulk solutions that are not smooth. In the first case, one can take a boundary metric which satisfies all the corner conditions at $t=0$, but then modify it in a smooth but nonanalytic way so that it becomes static. In the second case, one can insist on a static boundary metric everywhere and violate the corner conditions at $t=0$. This results in a null gravitational ``shock wave" of lower differentiability in the solution.

We close with some open questions. Of course, the main one is to confirm the existence of this nonlinear instability and understand it better. Here are some others:
\begin{enumerate}

\item Does global AdS in four and six dimensions exhibit this instability? (This requires breaking more symmetry.)
\item Does the AdS soliton exhibit this instability? (We have seen in four dimensions that the mass terms can act to slow down the growth, but they may not be sufficient to stop it.)
\item What does the nonlinear instability look like in odd dimensions with log terms in the initial data?
\item We have only considered vacuum solutions. How do matter fields (with analytic boundary data) affect this instability? 
\item In the context of holography, does this instability have any implications for the dual gauge theory?
\end{enumerate}

\section*{Acknowledgements}

It is a pleasure to thank Ted Jacobson and Stefan Hollands for discussions. This work was supported in part by NSF grants PHY-1801805 and PHY-1748958.  Use was made of computational facilities purchased with funds from the National Science Foundation (CNS-1725797) and administered by the Center for Scientific Computing (CSC). The CSC is supported by the California NanoSystems Institute and the Materials Research Science and Engineering Center (MRSEC; NSF DMR 1720256) at UC Santa Barbara.

\bibliographystyle{JHEP}
\bibliography{library}

\end{document}